\documentclass[aps,prb,reprint,amsmath,amssymb,superscriptaddress]{revtex4-2}

\usepackage{graphicx}
\usepackage{dcolumn}
\usepackage{bm}
\usepackage{amsmath}
\usepackage{lipsum}

\usepackage{mathrsfs}
\usepackage{mathtools}
\usepackage{xcolor}
\usepackage{orcidlink}
\usepackage[capitalise]{cleveref}
\usepackage{hyperref}
\usepackage{titlesec}
\usepackage{appendix}
\titleformat{\section}[runin]
  {\itshape}
  {} 
  {0.5em}
  {}

\titlespacing*{\section}
  {2pt}{2ex}{0em}

\newcommand{\setupappendixstyle}{%
  \titleformat{\section}[runin]
    {\itshape}
    {Appendix~\thesection.}
    {0.5em}
    {}%
  \renewcommand{\thesection}{\Alph{section}} 
}
\begin{document}

\preprint{APS/123-QED}

\title{Active Brownian particles in power-law viscoelastic media}

\author{D.~S. Quevedo\orcidlink{0000-0003-1583-4262}}
 \affiliation{Institute for Theoretical Physics, Utrecht University, Princetonplein 5, 3584CC Utrecht, The Netherlands}
\author{M. Conte\orcidlink{0009-0000-2632-6306}}
\affiliation{Soft Condensed Matter and Biophysics, Debye Institute for Nanomaterials Science, Utrecht University, Utrecht, The Netherlands}
\author{M. Dijkstra\orcidlink{0000-0002-9166-6478}}%
\affiliation{Soft Condensed Matter and Biophysics, Debye Institute for Nanomaterials Science, Utrecht University, Utrecht, The Netherlands}
\author{C. Morais Smith\orcidlink{0000-0002-4190-3893}}%
\affiliation{Institute for Theoretical Physics, Utrecht University, Princetonplein 5, 3584CC Utrecht, The Netherlands}

\date{\today}

\begin{abstract}
Many active particles are embedded in environments that exhibit viscoelastic properties. An important class of such media lacks a single characteristic relaxation timescale when subjected to a time-dependent stress. Rather, the stress response spans a broad continuum of timescales, a behavior naturally described by a scale-free, fractal-like power-law relaxation modulus. Using a generalization of the fractional Langevin equation, we investigate an active Brownian particle embedded in a power-law viscoelastic environment with translational and rotational dynamics governed by independent fractional orders. We solve the model analytically, develop a numerical scheme to validate the theoretical predictions, and provide tools that can be used in further studies. A rich variety of diffusion regimes emerges, which modify the intermediate-time behavior of the mean squared displacement. Notably, we find that the competition between translational and rotational contributions favors a superdiffusive persistence over the standard ballistic motion, and over-stretches its characteristic timescale, fundamentally altering the standard relation between persistence and propulsion in active matter.
\end{abstract}

\maketitle

Active Brownian particles are an important example of active matter, capable of converting internal energy into directed motion \cite{catesactivematter,active_part_intro1, active_part_intro2}. Active-matter systems were originally studied in the context of collective motion at both macroscopic and microscopic scales \cite{vicsek, vicsek2,collmotionZhaoxuan}. They subsequently became a central focus because of their ability to describe phenomena across vastly different length scales, from herds of large mammals to bacteria colonies, under minimal modeling assumptions \cite{animal_coll, bacteria_swimming}. Particular attention has been given to both biological and synthetic micro-swimmers, micron-sized particles that propel themselves through fluid environments, because of their relevance for the description of bacterial behavior and for the development of engineered self-propelled devices \cite{microswimmers, nanomotors}. The interplay between persistent active motion, thermal fluctuations, and environmental interactions, gives rise to a rich variety of emergent phenomena such as anomalous diffusion, clustering, and non-equilibrium phase transitions \cite{jerky, anom_coll, activephases, microtubes}.

Often, micro-swimmers are embedded in materials that deviate from the ideal behaviors described by Hooke’s law for elastic solids and Newton’s law for viscous fluids. Instead, such materials exhibit both elastic and viscous responses simultaneously and are therefore classified as viscoelastic \cite{viscoelastic_mat_book}. A useful way to characterize their viscoelastic properties is through a relaxation test, which measures the stress response to a time-dependent applied strain. This response is described by the so-called relaxation modulus, which is constant for ideal elastic solids and proportional to a Dirac delta function for purely viscous fluids. The motion of a particle embedded in such a medium can then be modeled using a generalized Langevin equation (GLE), where the memory kernel is related to the relaxation modulus via the generalized Stokes–Einstein relation (GSE) \cite{Mason2000, Mason1997}.

The form of the memory kernel plays a crucial role in determining the system's dynamical behavior. To date, GLEs with exponential memory kernels have been widely employed \cite{Narinder, sprenger2022active}, but a broad class of soft and biological materials exhibits a power-law relaxation modulus that decays as $\chi(t) \propto t^{-\gamma}$ \cite{frac_visc_for_power_law}. Using the GSE, the corresponding friction kernel scales as $K(t)\propto t^{-\alpha}$, with $\gamma = \alpha + 1$ \cite{manson1995}. Such power-law responses are characteristic of complex materials possessing a wide distribution of microstructural length and time scales. Important examples include biological tissues \cite{tumors}, gels \cite{silica_gel, colloid_gel, glutengels}, intra- and extracellular components \cite{actine, actine2, intracellular, lipidbilayer}, polymer networks \cite{polymers_frac, ploymers2, polymers3}, and soft crowded suspensions \cite{Mason1997, Yuki, crowded_cytoplasm}. Although understanding active motion in these environments is of considerable importance, it remains largely unexplored, with only a few exceptions \cite{joo2023viscoelastic, prev_work_nov}, all of which have been restricted to one-dimensional systems. 

Fractional calculus provides a natural framework for describing power-law relaxation moduli and memory kernels. Fractional operators encode scale-free, fractal-like memory, and therefore do not require the introduction of a characteristic timescale. Indeed, fractional models have proven highly successful in describing the rheological behavior of soft biological tissues exhibiting power-law scaling \cite{metzler_rehology, Hilfer2000, brain_liver, fract_in_bio}. Furthermore, they constitute the cornerstone for the study of anomalous diffusion behavior \cite{metzler2000random}, where the mean squared displacement (MSD) deviates from linear growth and instead scales as $t^{\alpha}$. The fractional generalization of the Langevin equation, first introduced by Lutz \cite{lutz2001fractional}, is particularly relevant in this context. It captures the dynamics of a particle subjected to a dissipative force of non-integer order and to long-range, time-correlated noise. The interplay between dissipation and noise gives rise to slow relaxation dynamics, leading to anomalous diffusion and, more recently, to the emergence of pre-thermal periodic time-ordered phases \cite{quevedo2024emergent, verstraten2021time}. 

In this work, we extend the fractional Langevin equation to model the motion of active Brownian particles, in which both translational and rotational degrees of freedom are influenced by the power-law viscoelastic properties of the surrounding medium. We solve the model analytically and develop numerical expressions suitable for further investigations. The interplay between the fractional orders and the noise acting on the spatial degrees of freedom gives rise to a rich spectrum of anomalous diffusion regimes that modifies the intermediate-time behavior of the MSD favoring superdiffusion over the standard $t^2$ during the persistence phase and preserving the long-time enhanced normal diffusion that characterizes active particles.

\section{Fractional active Langevin equation\textbf{---}}
We study the dynamics of an active Brownian particle embedded in a two-dimensional viscoelastic medium, described by its position vector $\boldsymbol{r}(t)=(x(t),y(t))$ and orientation angle $\phi(t)$. Both state variables evolve according to the following set of overdamped generalized Langevin equations,
\begin{equation}
\begin{aligned}[b]
    \int_{0}^tK_T(t-\tau)[\dot{\boldsymbol{r}}(\tau) - \upsilon\boldsymbol{\hat{n}}(\tau)]d\tau = & \theta_T\boldsymbol{\xi}_T(t), \\
    \int_{0}^tK_R(t-\tau)\dot{\phi}(\tau) d\tau = & \theta_R\xi_R(t),
\end{aligned}
    \label{eq:aGLe}
\end{equation}
where $\boldsymbol{\hat{n}}=(\cos\phi, \sin\phi)$ denotes the direction of self-propulsion with magnitude $\upsilon$, and $\boldsymbol{\xi}_T=(\xi_x, \xi_y)$ and $\xi_R$ are the translational and rotational stochastic forces, respectively, with strengths $\theta_{\{T,R\}}$. The time-delayed viscoelastic response of the medium is captured by a power-law memory kernel,
\begin{equation}
K_Q(t-t') = \frac{\eta_Q(t-t')^{-\alpha_Q}}{\Gamma(1-\alpha_Q)},
    \label{eq:K}
\end{equation}
where $\alpha_Q \in (0,1)$,  $\eta_{Q}$  is the corresponding viscous coefficient, and the subscript $Q = \{T,R\}$ denotes either the translational or rotational component. This choice corresponds to a power-law  relaxation modulus of the form $\chi(t)\propto t^{-\alpha-1}$, characteristic of scale-free viscoelastic media. Using the Caputo fractional derivative \cite{caputo1967linear},
\begin{equation}
    \prescript{C}{{t_0}}{D}^{\alpha_Q}_t\boldsymbol{f} = \int_{t_0}^t\frac{(t-\tau)^{-\alpha_Q}}{\Gamma(1-\alpha_Q)}\frac{d\boldsymbol{f}}{d\tau}d\tau,
    \label{eq:Caputo}
\end{equation}
Eqs.~\eqref{eq:aGLe} can be rewritten as a set of overdamped fractional Langevin equations,
\begin{equation}
\begin{aligned}[b]
    \eta_T\prescript{C}{{0}}{D}^{\alpha_T}_t[\boldsymbol{r}(t) - \upsilon\boldsymbol{\hat{N}}(t)] = & \theta_T \boldsymbol{\xi}_T(t), \\
    \eta_R\prescript{C}{{0}}{D}^{\alpha_R}_t\phi(t) = & \theta_R \xi_R(t),
\end{aligned}
    \label{eq:afLe}
\end{equation}
where the time derivative of $\boldsymbol{\hat{N}}$ defines the direction of self-propulsion $\boldsymbol{\hat{n}}=d\boldsymbol{\hat{N}}/dt$. In the limit $\alpha_Q\rightarrow 1$, the Caputo fractional derivative in Eq.~\eqref{eq:Caputo} reduces to an ordinary first-order time derivative, and Eqs.~\eqref{eq:afLe} recover the conventional overdamped active Langevin dynamics.

In particular, we model the stochastic forces as fractional Gaussian noises with zero mean, $\langle\boldsymbol{\xi}_T(t)\rangle=0$ and $\langle\xi_R(t)\rangle=0$, and power-law correlations for $t\neq t'$,
\begin{equation}
    \begin{aligned}[b]
\mathbb{C}_T & \equiv \theta_T^2\langle\boldsymbol{\xi}_T(t)\otimes\boldsymbol{\xi}_T(t')\rangle
=\mathbb{I}\theta_T^2H_T(2H_T-1)|t-t'|^{2H_T-2}, \\
C_R & \equiv \theta_R^2 \langle\xi_R(t)\xi_R(t')\rangle
=\theta_R^2 H_R(2H_R-1)|t-t'|^{2H_R-2},
    \end{aligned}
    \label{eq:correlations}
\end{equation}
where $\mathbb{I}$ denotes the $2\times2$ identity matrix. Isotropy of the medium implies a single, constant Hurst exponent $H_T$ for both translational degrees of freedom, while the rotational noise is characterised by a (generally  distinct from the translational one) Hurst exponent $H_R$.

For simplicity, we assume that the memory kernels are related to the correlation functions of the noise through the fluctuation-dissipation relations of the second kind,
\begin{equation}
\begin{aligned}[b]
    \mathbb{C}_T(|t-t'|) &= k_BT\mathbb{I}K_T(t-t'),\\
    C_R(|t-t'|) &  = k_BTK_R(t-t'),
\end{aligned}
    \label{eq:fluctuation-dissipation}
\end{equation}
where $k_B$ is the Boltzmann's constant and $T$ the temperature. 
As a consequence, the order of the fractional derivative and the Hurst exponent are related by $\alpha_Q=2-2H_Q$, and the noise amplitude is given by $\theta_Q^2=\eta_Q k_BT/[\Gamma(1-\alpha_Q)H_Q(2H_Q-1)]$.

The solutions of Eqs.~\eqref{eq:afLe} can be obtained using Laplace transform techniques (see Refs.~\cite{risken1989fokker,kwok2018langevin}). By Laplace transforming Eqs.~\eqref{eq:afLe} and inverting back, we obtain
\vspace{-0.7\baselineskip}
\begin{equation}
    \begin{aligned}[b]
      \boldsymbol{r}(t)& =\boldsymbol{r}_0 + \theta_T(\boldsymbol{\xi}_T*G_T)(t) - \upsilon\int_0^t\boldsymbol{\hat{n}}(\tau)d\tau, \\
      \phi(t)   & =\phi_0 + \theta_R(\xi_R*G_R)(t),
    \end{aligned}
    \label{eq:solutions}
\end{equation}
where $\boldsymbol{r}_0=\boldsymbol{r}(t=0)$, $\phi_0=\phi(t=0)$, $*$ is the convolution operator, and $G_Q(t)$ is the resolvent function defined as the inverse Laplace transform of $\tilde{G}_Q(s) = 1/[s\tilde{K}_Q(s)]$. Recalling the power-law memory kernel in Eq.~\eqref{eq:K}, whose Laplace transform is $\tilde{K}_Q(s)=\eta_Qs^{\alpha_Q-1}$, the resolvent function becomes
\vspace{-0.7\baselineskip}
\begin{equation}
    \begin{aligned}[b]
      G_Q(t)=\mathcal{L}^{-1}\{\tilde{G}_Q(s);t\} = \frac{t^{\alpha_Q-1}}{\eta_Q\Gamma(\alpha_Q)}.
    \end{aligned}
    \label{eq:G}  
\end{equation}

\section{Numerical solutions of the fractional active Langevin equation\textbf{---}}

The particle's inertia combined with the slow relaxation dynamics typical of the fractional Langevin equations can give rise to  transient pre-thermal effects for small values of $\alpha_Q$, manifested as periodic oscillations in the velocity autocorrelation and the MSD \cite{quevedo2024emergent}. This feature poses challenges for simulating fractional overdamped dynamics because numerical algorithms developed for underdamped cases \cite{guo2013numerics,quevedo2024emergent} converge slowly to the overdamped limit. To overcome this problem, we use the L1 discretization scheme of the Caputo operator \cite{L1_Caputo_discretization} (see App.~\ref{app:numerics} of the End Matter), allowing us to tailor the recursive discretized expression for the overdamped rotational and the active translational equations,
\vspace{-0.7\baselineskip}
\begin{equation}
\begin{aligned}[b]
\phi(t_n) & = \phi(t_{n-1})-\sum_{j=1}^{n-1}c_{n-j}\left[ \phi(t_j)- \phi(t_{j-1})\right] \\ 
&+\frac{\Gamma(2-\alpha_R)}{h^{-\alpha_R}}\theta_R\xi_R(t_n), \\
\end{aligned}\label{eq:L1_discr_phi}
\end{equation}
\begin{equation}
\begin{aligned}[b]
&\boldsymbol{r}(t_n) 
= \boldsymbol{r}(t_{n-1}) +\upsilon \left[ \boldsymbol{\hat{N}}(t_n) - \boldsymbol{\hat{N}}(t_{n-1}) \right]\\
&+ \sum_{j=1}^{n-1} c_{n-j} \left\{ \upsilon\left[ \boldsymbol{\hat{N}}(t_j) - \boldsymbol{\hat{N}}(t_{j-1}) \right] - \left( \boldsymbol{r}(t_j) - \boldsymbol{r}(t_{j-1}) \right) \right\} \\
& + \frac{\Gamma(2-\alpha_T)}{h^{-\alpha_T}}\theta_T\boldsymbol{\xi}_T(t_n),
\end{aligned}\label{eq:L1_discr_r}
\end{equation}
where $c_k = (k+1)^{1-\alpha_Q} - k^{1-\alpha_Q}$, $\boldsymbol{\hat{N}}(t_j) - \boldsymbol{\hat{N}}(t_{j-1}) \approx h[\cos\phi(t_j),\sin\phi(t_j)]$, and the algorithm achieves a local precision of  $O(h^{2-\alpha_Q})$\cite{li2015numerical,L1_Caputo_discretization}.

We calculated numerical solutions for the angular and translational components ($\phi(t_n), x(t_n), y(t_n)$) using Eqs.~\eqref{eq:L1_discr_phi} and ~\eqref{eq:L1_discr_r} with uniformly distributed initial angle $\phi_0 \sim U[0,\pi]$ and $x_0 = y_0 = 0$. The rotational and translational noises were generated from fractional Brownian motion trajectories using the Python library fBm0.3.0  \cite{flynn2019fbm}, and then differentiated to obtain the fractional Gaussian noises $\theta_R\xi_R(t_n)= [B_{H_R}(t_n) - B_{H_R}(t_{n-1})]h^{-1}$ and $\theta_T\boldsymbol{\xi}_T(t_n)= [\boldsymbol{B}_{H_T}(t_n) - \boldsymbol{B}_{H_T}(t_{n-1})]h^{-1}$. Numerical MSDs were calculated directely as  $(\phi(t_n) - \phi_0)^2$ and $(\boldsymbol{r}(t_n) - \boldsymbol{r}_0)^2$, averaged over 3000 independent trajectories. We 
defined the characteristic timescale of the rotational component as $\tau_R = [\Gamma(1+\alpha_R)D_R/2]^{-1/\alpha_R} = [\eta_R/k_BT]^{1/\alpha_R}$
, where $D_R=2k_BT/[\eta_R\Gamma(1+\alpha_R)]$ is the rotational diffusion coefficient. For all the numerics, we computed $95\%$ normal confidence intervals.

\begin{figure}[ht]
    \centering
    \includegraphics[width=0.75\columnwidth]{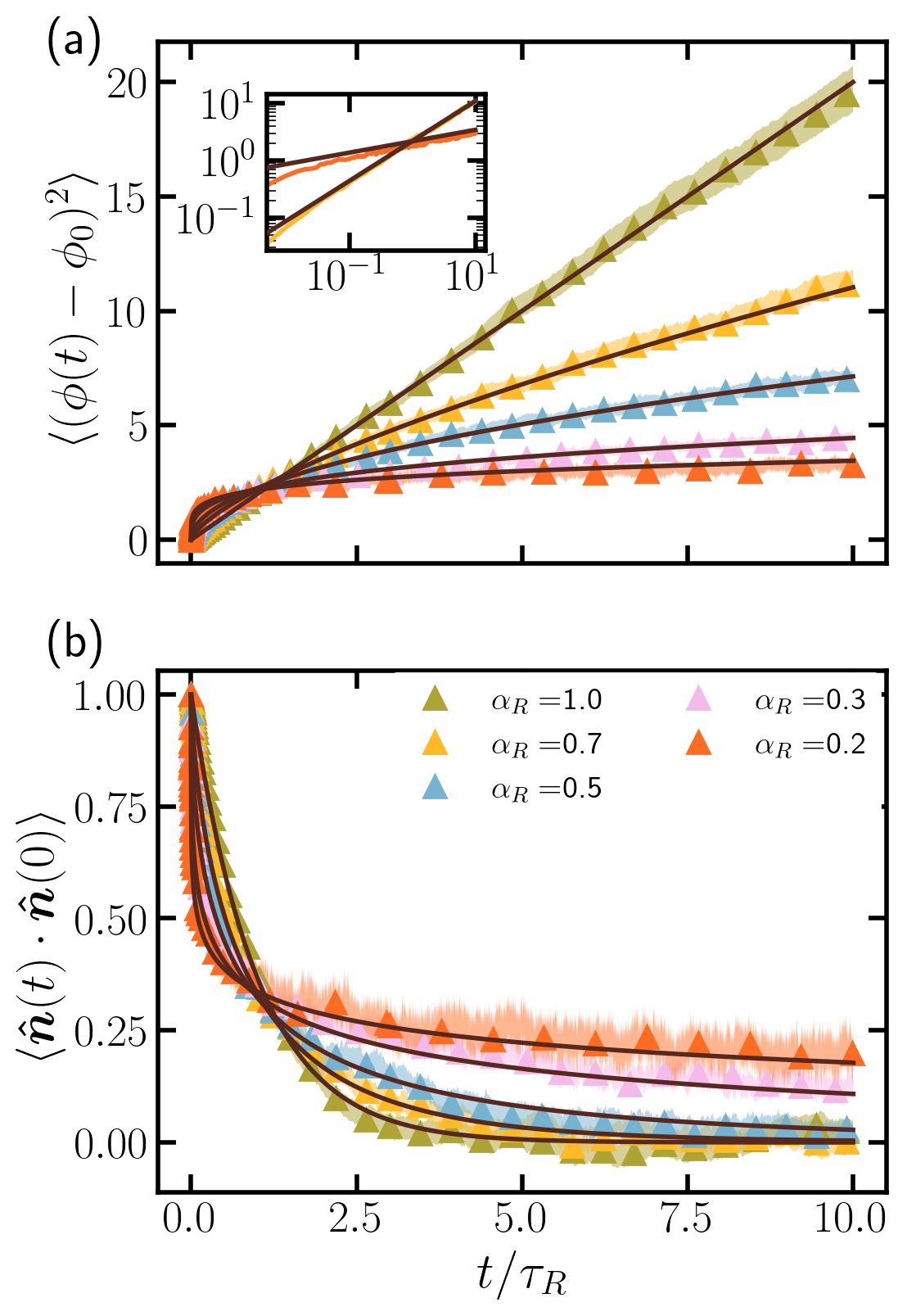}
    \caption{Dynamics of the angular component of an active Brownian particle for different fractional orders $\alpha_R$. Analytical expressions are shown as solid brown lines, while colored markers represent sampled points from numerical simulations with $95\%$ confidence bands. (a) Angular MSD in units of $\textrm{rad}^2$. The inset illustrates the thermalization of the numerical solutions toward the overdamped limit for $\alpha_R=0.2$ (orange) and $\alpha_R=0.7$ (yellow), using a time step $h=0.005$.
    (b) Autocorrelation function of the self-propulsion direction. All curves are normalized by the rotational timescale $\tau_R = (\eta_R/k_BT)^{1/\alpha_R}$.}
    \label{fig:phi_stats}
\end{figure}

In \cref{fig:phi_stats}(a), we present the numerical calculation of the angular MSD and compare it with the analytical result $\langle (\phi(t)-\phi_0)^2\rangle = D_Rt^{\alpha_R}$ derived in the App.~\ref{app:fLe} of the End Matter. Analytical expressions are shown as brown solid lines, while colored markers represent sampled points from the numerical solutions, and the shaded areas indicate the $95\%$ confidence intervals. The same scaling holds for the passive translational motion ($\upsilon=0$), with an additional factor of 2 accounting for the contributions of both $x$ and $y$. The thermalization behavior of the numerical algorithm is analyzed in the inset of \cref{fig:phi_stats}(a) for fixed $h=0.005$ and $\alpha_R=\{0.2, 0.7\}$. For $\alpha_R=0.2$ (orange), the convergence to the overdamped limit occurs on a time scale $\mathcal{O}[10t/\tau_R]$, whereas for $\alpha_R=0.7$ (yellow), the convergence is much faster, $\mathcal{O}[10^{-1}t/\tau_R]$. To improve accuracy and speed-up thermalization, we therefore set the time step at $h=5 \times 10^{-5}$ for $\alpha_R=0.2$, $h=0.001$ for $\alpha_R=0.3$, and kept $h=0.005$ for $\alpha_R>0.3$, in all subsequent numerical calculations.
\section{Autocorrelation of the self-propulsion direction\textbf{---}}
A common assumption for GLEs driven by Gaussian noise is that their solutions are linear combinations of Gaussian processes,  so the process itself is asymptotically Gaussian \cite{wang1992long,porra1996generalized}. Although the rigorous construction of the corresponding Fokker--Planck equation for a non-Markovian process remains an open problem \cite{calvo2008path,hanggi1978correlation,khan2005derivation,quevedo2025path}, and is generally non-Gaussian (except in the case of fractional Brownian motion \cite{calvo2008path}), this Gaussian approximation allows us to simplify the analytical calculations and to systematically study the statistical properties of the self-propulsion direction. Hence, we assume that the angular displacement follows a Gaussian distribution with zero-mean $\mu_{\phi} = 0$ and variance $\sigma_{\phi}^2 = \langle (\phi - \phi_0)^2 \rangle + |\phi_0|^2$. The autocorrelation of the self-propulsion direction then reads
 \vspace{-0.7\baselineskip}
\begin{equation}
    \begin{aligned}[b]
        \langle\boldsymbol{\hat{n}}(t)\cdot\boldsymbol{\hat{n}}(t')\rangle = \langle\cos[\Delta\phi(t)]\rangle= 
     e^{-\frac{D_{R} |t-t'|^{\alpha_R}}{2}},
    \label{eq:correlation_rot}
    \end{aligned}
\end{equation}
which stems from the standard result for the cosine of a Gaussian variable $X$ with mean $\mu_X$ and variance $\sigma_X^2$, $\langle\cos(X)\rangle=\Re{\langle\exp(iX)\rangle}=\cos(\mu_X)\exp(-\sigma_X^2/2)$ \cite{risken1989fokker,ghosh2015communication}.

In \cref{fig:phi_stats}(b), we present numerical results for the autocorrelation of the self-propulsion direction and compare them with Eq.~\eqref{eq:correlation_rot} for $t'=0$ and various values of $\alpha_R$. The area under the autocorrelation curves corresponds to the persistence time, which is the average time the particle maintains its orientation. Using the change of variables $s=D_Rt^{\alpha_R}/2$, it can be calculated by integrating Eq.~\eqref{eq:correlation_rot},
$\tau_p = \int_0^\infty \langle\boldsymbol{\hat{n}}(t)\cdot\boldsymbol{\hat{n}}(0)\rangle dt = (2/D_R)^{1/\alpha_R}\Gamma(1+1/\alpha_R)$.
For $\alpha_R=1$, it retrieves the standard persistence time for conventional active Brownian particles $\tau_p=2/D_R$, where the orientation decays exponentially. For $\alpha_R<1$, however, memory effects induced by the fractional operator stretch the decorrelation time. Consequently, decreasing  $\alpha_R$ increases the persistence of the orientation, which diverges in the limit $\lim_{\alpha_R\to0}\tau_P \rightarrow\infty$. Furthermore, we observed two distinct trends: for $\alpha_R\geq0.5$, the persistence time decays rapidly toward the lower limit $\alpha_R=1$, while for  $\alpha_R<0.5$, the decorrelation time is strongly stretched.

In Ref.~\cite{gomez2020active}, active Brownian particles were modeled with a rotational component described by linear friction and driven by fractional Brownian motion. In that case, the persistence time depends on the Hurst exponent $0<H_R<1$ as $(2/D_H)^{1/2H_R}$, where $D_H$ is a Hurst-dependent rotational diffusion coefficient. Smaller values of $H_R$ increase the persistence time and reinforce long-range correlations, while $H_R=1/2$ reduces to the conventional active Brownian particle behavior. In contrast, the order of the fractional derivative and the rotational Hurst exponent in Eqs.\eqref{eq:aGLe} are coupled via $\alpha_R=2-2H_R$, with $1/2 \leq H_R < 1$, as imposed by the fluctuation-dissipation theorem. This coupling implies that higher $H_R$ values strengthen long-range correlations and extends the persistence time, since our model is defined only within the persistent regime of fractional Brownian motion.
\begin{widetext}
\section{Active diffusion\textbf{---}}
The total MSD of an active particle can be calculated from the solutions given by Eq.~\eqref{eq:solutions}. Using the results derived in App.~\ref{app:fLe} and App.~\ref{app:msd_act_rot} of the End Matter, we obtain
\begin{equation}
\begin{aligned}[b]
\langle (\boldsymbol{r}-\boldsymbol{r}_0)^2 \rangle
& = \int_0^td\tau_1G_T(\tau_1)\int_0^td\tau_2G_T(\tau_2)Tr[\mathbb{C}_T(|\tau_1-\tau_2|)] + \upsilon^2 \int_0^td\tau_1\int_0^td\tau_2\langle\boldsymbol{\hat{n}}(\tau_1)\cdot\boldsymbol{\hat{n}}(\tau_2)\rangle \\
& = 2 D_Tt^{\alpha_T} + \frac{2\upsilon^2}{\alpha_R(D_{R}/2)^{\frac{2}{\alpha_R}}}\left[ \gamma\left(\frac{1}{\alpha_R},\frac{D_{R}}{2}t^{\alpha_R}\right)\left(\frac{D_{R}}{2}\right)^{\frac{1}{\alpha_R}}t- \gamma\left(\frac{2}{\alpha_R},\frac{D_{R}}{2}t^{\alpha_R}\right)\right],
    \label{eq:MSD_active}    
\end{aligned}
\end{equation}
where $D_T$ denotes the translational diffusion coefficient and $\gamma(a,z)$ is the lower incomplete Gamma function. 
\begin{figure}[ht]
    \centering
    \includegraphics[width=0.85\columnwidth]{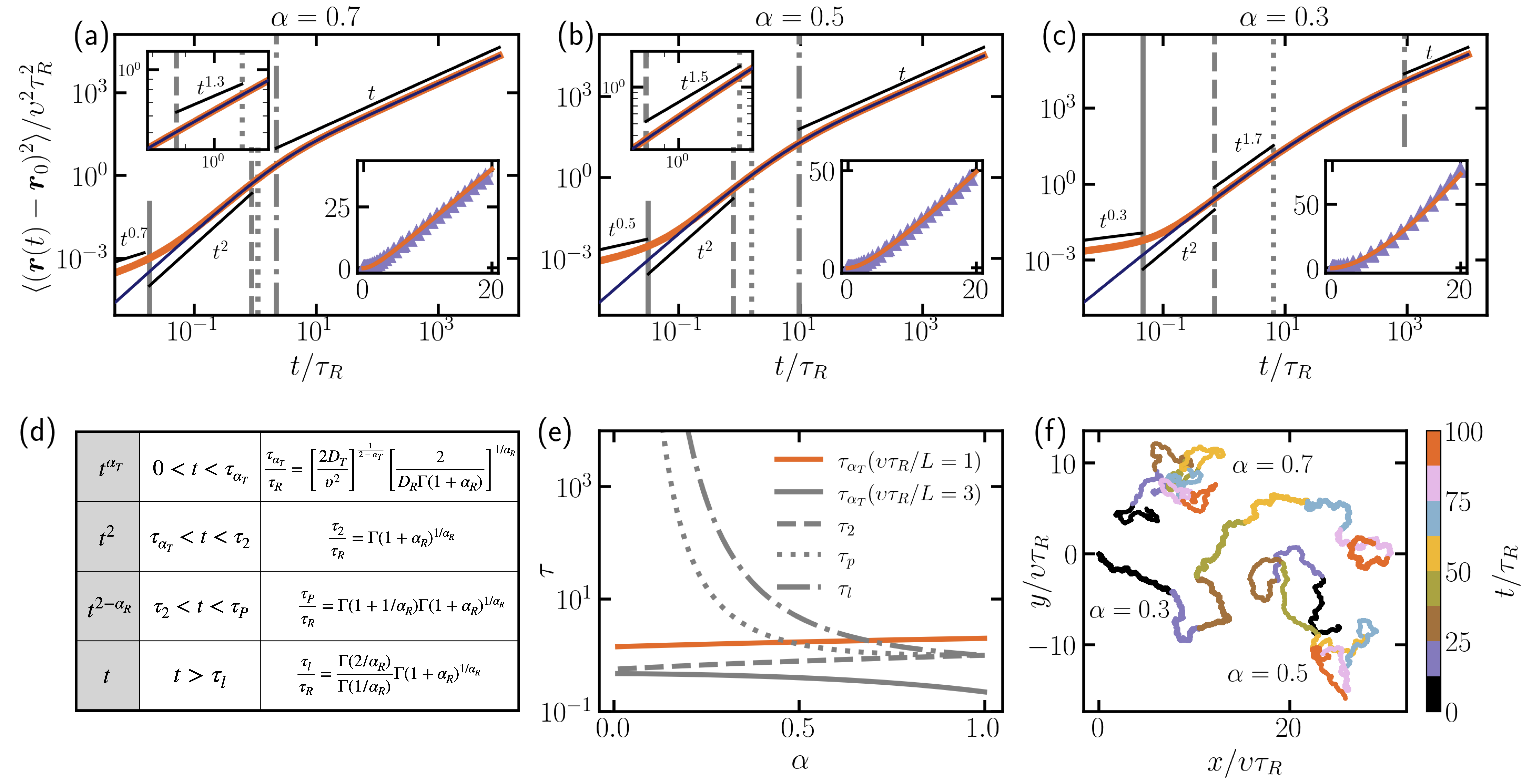}
    \caption{Active diffusion regimes for $\alpha\equiv\alpha_R=\alpha_T$. (a)-(c) MSD for $\upsilon\tau_R/L=20$ (solid orange line) and $\upsilon\tau_R/L=\infty$ (solid navy) for $\alpha={0.7,0.5,0.3}$, respectively. Bottom-right insets show the comparison between analytical and numerical results, while top-left insets present  a zoom into the $t^{2-\alpha_R}$ superdiffusive regime. Crossover times between diffusion regimes are indicated by grey lines using the convention: $\tau_{\alpha_T}$-solid, $\tau_{2}$-dashed, $\tau_{P}$-dotted and  $\tau_l$-dotted-dashed. (d) Summary of the diffusion regimes. (e) Crossover times  as a function of $\alpha$. (f) Representative particle trajectories for each $\alpha$. All times-scales are in units of $\tau_R$, and all spatial coordinates are normalized by the characteristic length $\upsilon\tau_R$.}
    \label{fig:phases}
\end{figure}
\end{widetext}

When $\alpha_R=\alpha_T=1$, Eq.~\eqref{eq:MSD_active} recovers the conventional result for active Brownian particles, where the MSD exhibits an initial short-time diffusive regime, followed by an active persistent scaling behavior $t^2$, and  finally a crossover to long-time normal diffusion $t$ with an enhanced diffusion coefficient \cite{Volpe2014}. Additionally, Eq.~\eqref{eq:MSD_active} reproduces the results of Ref.~\cite{gomez2020active} when imposing $\alpha_R=2H_R$,  corresponding to a fractional rotational noise with linear dissipation. However, when the fractional orders differ from one and satisfy the thermal condition $\alpha_Q=2-2H_Q$, a rich variety of dynamical regimes emerges. At short times, a competition between  translational subdiffusion and the active persistent phase occurs, with a crossover time $\tau_{\alpha_T} = (2D_T/\upsilon^2)^{1/(2-\alpha_T)}$ marking the transition from $t^{\alpha_T}$ to $t^2$. Within the persistent phase, the MSD  exhibits a second transition from ballistic $t^2$ to sub-ballistic superdiffusive $t^{2-\alpha_R}$. From the series expansion of the active MSD (see App.~\ref{app:msd_act_rot} of the End Matter), the characteristic crossover time is $\tau_2=(2/D_R)^{1/\alpha_R}$, since for small times $D_Rt^{\alpha_R}/2\ll1$, the leading contribution is $\upsilon t^2$. At long times, the system relaxes to normal diffusion with the long-time diffusion coefficient,
$D_l = 2\nu^2\Gamma(1+1/\alpha_R)/(D_R/2)^{1/\alpha_R}$, and the corresponding linear regime is reached for $t>>\tau_l$, where $\tau_l=(2/D_R)^{1/\alpha_R}\Gamma(2/\alpha_R)/\Gamma(1/\alpha_R)$ is the characteristic timescale at which the first incomplete gamma function dominates over the second one in Eq.\eqref{eq:MSD_active}.

In \cref{fig:phases}, we illustrate the diffusion regimes for fixed $\alpha\equiv \alpha_R=\alpha_T$, with values $\alpha=\{0.7,0.5,0.3\}$. Panels (a)-(c) present the analytical MSD from Eq.~\eqref{eq:MSD_active}, with $\upsilon\tau_R/L=20$ in orange solid lines and $\upsilon\tau_R/L=\infty$ in solid navy, where $L$ denotes a unit length. The crossover times between regimes are represented by grey lines following the convention: $\tau_{\alpha_T}$-solid, $\tau_{2}$-dashed, $\tau_{P}$-dotted and  $\tau_l$-dotted-dashed. The case $\upsilon\tau_R/L=\infty$ is effectively reached when $\upsilon^2\gg 2D_T$, where the short-time behavior is fully dominated by the $t^2$ active persistent phase, and  the initial translational subdiffusion is negligible. In contrast, $\upsilon\tau_R/L=20$ still allows for a brief subdiffusive regime at very early times. The comparison between analytical and numerical results is shown in the bottom-right insets, while the top-left insets provide a zoom into the $t^{2-\alpha_R}$ super-diffusive regime. The sequence of MSD regimes is summarized in \cref{fig:phases}(d).
\begin{figure}[ht]
    \centering
    \includegraphics[width=0.75\columnwidth]{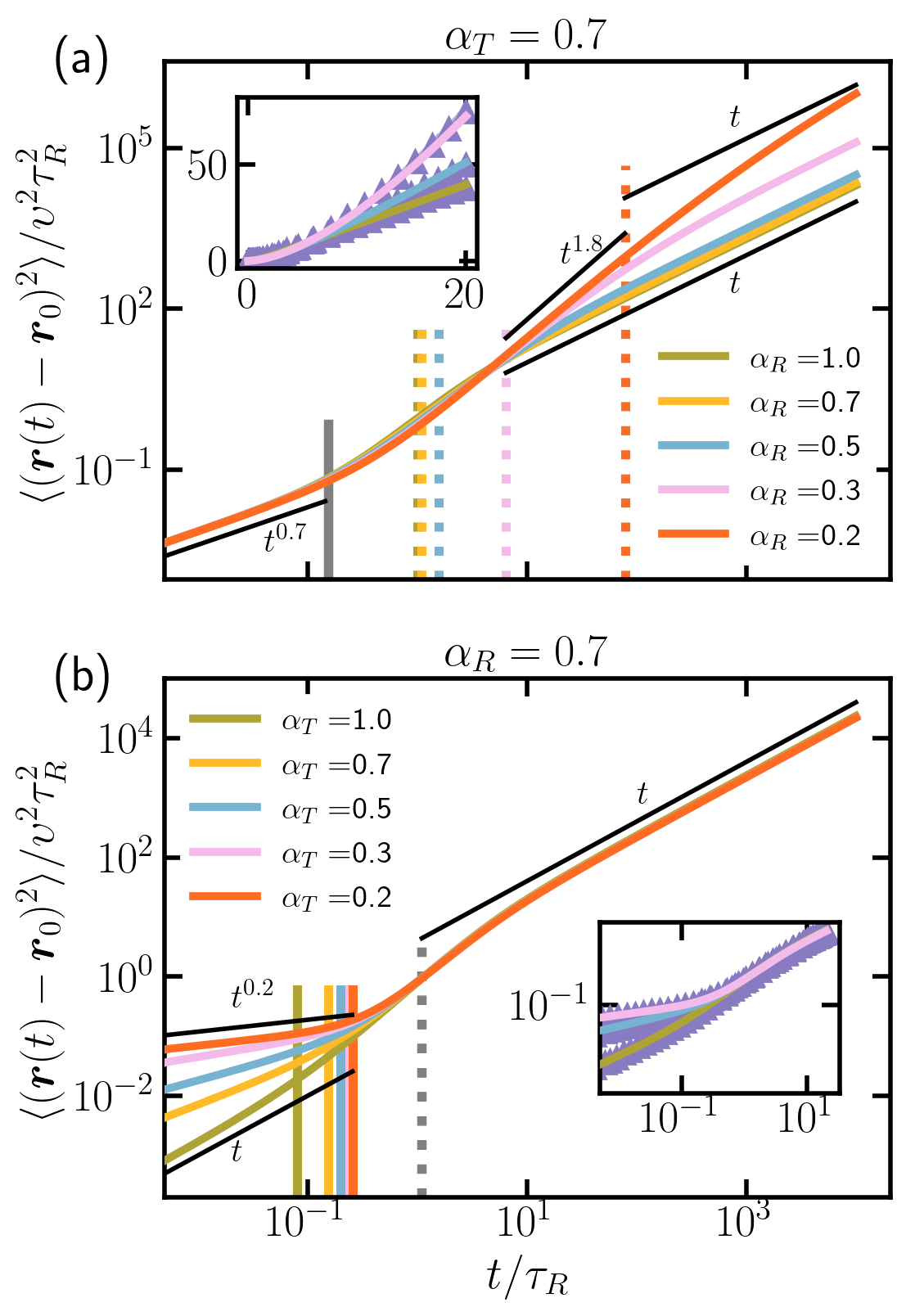}
    \caption{Diffusion regimes for decoupled translational and rotational orders. Analytical MSDs are presented in solid colored lines while numerical results are shown in purple in the figure insets. The relevant timescales $\tau_{\alpha_T}$ and $\tau_p$ are represented, respectively, in vertical solid and dotted lines, following the color convention of the MSD. (a) MSD for fixed $\alpha_T=0.7$. (b) MSD for  $\alpha_R=0.7$. Times are normalized by $\tau_R$ and spatial variables by the characteristic length $\upsilon\tau_R$.}
    \label{fig:diffarat}
\end{figure}

When $\alpha > 0.5$, the timescales $\tau_2, \tau_p, \tau_l$ approach each other and collapse at the Newtonian limit $\alpha =1$. Furthermore, for small $\upsilon\tau_R/L$, the timescale $\tau_{\alpha_T}$ partially dominates over $\tau_p$ and $\tau_l$ for $\alpha \geq0.5$ and fully suppresses the $t^2$ dynamics across the entire $\alpha$ range, as demonstrated by the orange line in \cref{fig:phases}(e). For $\alpha < 0.5$, both the superdiffusive regime and the approach to normal diffusion are over-stretched, diverging in the limit $\alpha \to0$, and favoring superdiffusion over ballistic behavior. Using Stirling's approximation for the Gamma function $\Gamma(z)=\sqrt{2\pi/z}(z/\text{e})^z[1+\mathcal{O}(1/z)]$, we analyze the divergent behavior of the characteristic timescales. For $z=1+1/\alpha $ with $1/\alpha  \gg 1$, the persistence time is approximately $\tau_P/\tau_R=\Gamma(1+1/\alpha )\Gamma(1+\alpha )^{1/\alpha } \approx \sqrt{2\pi/\alpha }(1/\alpha\text{e})^{1/\alpha }$, $\tau_R$ normalizes the timescale to remove the dependence on $D_R$, and $\Gamma(1+\alpha )^{1/\alpha }\approx1$. Similarly, the long-time diffusion scale is $\tau_l/\tau_R=\Gamma(2/\alpha )\Gamma(1+\alpha )^{1/\alpha }/\Gamma(1/\alpha )\approx\sqrt{2}(4/\alpha\text{e})^{1/\alpha}$, so that  $\tau_l/\tau_p\approx\sqrt{\alpha /\pi}4^{1/\alpha }$. This shows that the effective diffusion time $\tau_l$ grows exponentially faster than the persistence time $\tau_p$ in the limit $1/\alpha \gg 1$. The dependence of the timescales on $\alpha$ is shown in \cref{fig:phases}(e), while \cref{fig:phases}(f) shows sample trajectories for each $\alpha$. A marked difference is observed for $\alpha=0.3$, where the over-stretching of $\tau_p$ causes superdiffusive motion extending up to $t/\tau_R\approx100$. 

Finally, we analyze the effect of decoupling $\alpha_R$ and $\alpha_T$. In \cref{fig:diffarat}, we illustrate this by fixing either $\alpha_T=0.7$ or $\alpha_R=0.7$. The analysis of time scales presented above holds when $\alpha_R\neq\alpha_T$, as there is no cross-dependence between time scales and fractional orders. When $\alpha_T$ is fixed and $\alpha_R$ is varied [\cref{fig:diffarat}(a)], the short-time dynamics is dominated by $t^{\alpha_T}$ up to the crossover time $\tau_{\alpha_T}$, while the persistent regime exhibits an initial $t^2$ ballistic behavior, followed by a super-diffusive spectrum between $t$ and $t^2$, and finally relaxes to linear diffusion at times determined by the dependence of $\tau_l$ on $\alpha_R$.  On the other hand, varying $\alpha_T$ only affects the short-time dynamics, see \cref{fig:diffarat}(b). As $\alpha_T$ decreases, the persistent phase becomes shorter due to the enhanced  memory effects associated with the dominant elastic properties of the medium, while the long-time diffusion remains governed by the fixed $\alpha_R$. 

\section{Conclusion\textbf{---}} Our results provide new insights into the dynamics of active particles in power-law viscoelastic materials, a class of media that is highly relevant in soft and biological matter. We uncovered a rich variety of diffusive, subdiffusive, and superdiffusive regimes and identified the corresponding characteristic length and time scales, revealing a competition between translational and rotational contributions. This competition results in a persistent regime that favors an over-streched superdiffusive motion over a ballistic one, effectively questioning the relation between persistance and propuslion in active matter. The analytical and numerical framework developed here can be extended to investigate more complex active behaviors, such as collective phases or optimized motion strategies. Moreover, we anticipate that our findings will contribute to the understanding of experimental observations of active particles in diverse media ranging from gels \cite{silica_gel, colloid_gel, glutengels}, polymer networks \cite{polymers_frac, ploymers2, polymers3} and crowded environments \cite{Mason1997, Yuki, crowded_cytoplasm}.

\bibliography{apssamp}

\begin{widetext}
\begin{center}
\textbf{End Matter}
\end{center}
\end{widetext}
\appendix
\setupappendixstyle
\section{\label{app:numerics} L1 discretization of the fractional active Langevin equations\textbf{---}}
On a uniform time grid $t_j=k*h$ with $h$ the step size and $k\in[0,\dots, n]$, the integral over the time interval in \cref{eq:Caputo} can be split into the sum of $n$ contributions,
\begin{equation*}
    \prescript{C}{{t_0}}{D}^{\alpha_Q}_t q = \frac{1}{\Gamma(1-\alpha_Q)} \sum_{j=1}^{n}\int_{t_{j-1}}^{tj}(t_n-\tau)^{-\alpha_Q}q'(\tau)d\tau.
\end{equation*}
The first-order derivative of $q$ can be approximated using a finite difference scheme $hq'(\tau)\approx q(t_j)-q(t_{j-1})$, leading to
\begin{multline*}
    \int_{t_{j-1}}^{tj}(t-\tau)^{-\alpha_Q}\frac{dq}{d\tau}d\tau \approx \\ \frac{q(t_j)-q(t_{j-1})}{h}\int_{t_{j-1}}^{tj}(t_n-\tau)^{-\alpha_Q}d\tau.
\end{multline*}
The kernel integral can be exactly evaluated to
\begin{equation*}
    \int_{t_{j-1}}^{tj}(t-\tau)^{-\alpha_Q}d\tau = \frac{(t_n-t_{j-1})^{1-\alpha_Q}-(t_n-t_j)^{1-\alpha_Q}}{(1-\alpha_Q)},
\end{equation*}
where $t_n -t_{j-1}=(n-j+1)h$ and $t_n -t_{j}=(n-j)h$. Factoring out the time step $h$, we get the L1 Caputo discretization
\begin{equation}
\prescript{C}{{t_0}}{D}^{\alpha_Q}_t q = \frac{h^{-\alpha_Q}}{\Gamma(2-\alpha_Q)} \sum_{j=1}^{n} c_{n-j}\left[ q(t_j) - q(t_{j-1})\right],
\label{eq:L1_Caputo_discr_q}
\end{equation}
with $c_k = (k+1)^{1-\alpha_Q} - k^{1-\alpha_Q}$.

Applying \cref{eq:L1_Caputo_discr_q} to the rotational and translational components of the fractional active Langevin equation, we obtain the recursive relations in Eqs.~\eqref{eq:L1_discr_phi} and ~\eqref{eq:L1_discr_r}. In particular, the increments of the  self-propulsion direction can be further simplified by assuming a right-endpoint approximation,  
\begin{equation}
\begin{aligned}
\boldsymbol{\hat{N}}(t_j) - \boldsymbol{\hat{N}}(t_{j-1}) & = \int_{t_{j-1}}^{t_j} [\cos{\phi(t)},\sin{\phi(t)}]dt \\
& \approx h[\cos\phi(t_j),\sin\phi(t_j)].
\end{aligned}
\label{right_endpoint_approx}
\end{equation}

\begin{widetext}
\section{\label{app:fLe}Passive under- and over-damped dynamics\textbf{---}}
When $\upsilon\tau_R/L=0$, the set of fractional Eqs.~\eqref{eq:afLe} has the same form for both spatial and angular degrees of freedom. In this case, we can generically write the overdamped fractional Langevin equation,
\begin{equation}
    \eta_Q\prescript{C}{{0}}{D}^{\alpha_Q}_t q(t) = \theta_Q \xi_q(t),
    \label{eq:fLe_Q}
\end{equation}
with $q=\{x,y,\phi\}$ representing a general coordinate for space and angle, and $\xi_q=\{\xi_x,\xi_y,\xi_R\}$ the corresponding noise satisfying Eq.~\eqref{eq:correlations}. The general solution to Eq.\eqref{eq:fLe_Q} is given by $q(t) =q_0 + \theta_Q(\xi_q*G_Q)(t)$. Therefore, the passive MSD is
\begin{equation}
\begin{aligned}[b]
    \langle (q - q_0)^2 \rangle 
    & = \int_0^td\tau_1G_Q(\tau_1)\int_0^td\tau_2G_Q(\tau_2)C_q(|\tau_1 - \tau_2|) \\
    & = 2\int_0^td\tau_1G_Q(\tau_1)\int_0^{\tau_1}d\tau_2G_Q(\tau_2)C_q(\tau_1 - \tau_2 )\\
    & = \frac{2\theta_q^2H_Q(2H_Q-1)}{\eta_Q^2\Gamma^2(\alpha_Q)}\int_0^td\tau_1\;\tau_1^{\alpha_Q-1}\int_0^{\tau_1}d\tau_2\;\tau_2^{\alpha_Q-1}(\tau_1 - \tau_2 )^{-\alpha_Q}, \\
\end{aligned}
    \label{eq:MSD_passive_1}
\end{equation}
where the second line relies on splitting the squared integration domain into two equal triangular areas at $\tau_1=\tau_2$ and exchanging the variables $\tau_1\leftrightarrow\tau_2$. The second integral can be simplified by converting it into an Euler Beta function $\mathbf{B}(a,b)=\int_0^1 z^{a-1}(1-z)^{b-1}$, using the change of variables $\tau_2=\tau_1z$ together with the properties $\mathbf{B}(a,b)=\Gamma(a)\Gamma(b)/\Gamma(a+b)$ and $z\Gamma(z)=\Gamma(z+1)$,
\begin{equation}
\begin{aligned}[b]
    \langle (q - q_0)^2 \rangle 
    & = \frac{2\theta_q^2 H_Q(2H_Q-1)}{\eta_Q^2\Gamma^2(\alpha_Q)}\int_0^td\tau_1\;\tau_1^{\alpha_Q-1}\int_0^1dz\;z^{\alpha_Q-1}(1 - z )^{-\alpha_Q} \\ 
    & = \frac{2\theta_q^2 H_Q(2H_Q-1)\mathbf{B}(\alpha_Q,1-\alpha_Q)}{\eta_Q^2\Gamma^2(\alpha_Q)}\int_0^td\tau_1\;\tau_1^{{\alpha_Q-1}} \\
    & = \frac{2\theta_q^2 H_Q(2H_Q-1)\Gamma(1-\alpha_Q)}{\eta_Q^2\Gamma(\alpha_Q+1)}t^{{\alpha_Q}} \\
    & = D_{Q} t^{\alpha_Q},
\end{aligned}
    \label{eq:MSD_passive_2}
\end{equation}
where we replaced $\theta_Q^2=\eta_Tk_BT/[\Gamma(1-\alpha_Q)H_R(2H_R-1)]$ and defined the diffusion coefficient $D_{Q} = 2k_BT/[\eta_Q\Gamma(1+\alpha_Q)]$.

The underdamped counterpart of Eq.~\eqref{eq:fLe_Q} is the so-called fractional Langevin equation \cite{lutz2001fractional},
\begin{equation}
    m\ddot{q}(t) + \eta_Q\prescript{C}{{0}}{D}^{\alpha_Q}_t q(t) = \theta_Q \xi_q(t).
    \label{eq:fLe_Qu}
\end{equation}
The resolvent function of this equation is the more general result written in terms of the Mittag-Leffler function $E_{a,b}(z) = \sum_{k=0}^\infty z^k/\Gamma(ak+b)$ \cite{mainardi2000mittag},
\vspace{-0.7\baselineskip} 
\[
G_{u} = t\,E_{2-\alpha_Q,2}\!\left(\frac{\eta_Q}{m} t^{2-\alpha_Q}\right),
\]
where the typical relaxation timescale is $\tau_u =(m/\eta_Q)^{1/2-\alpha_Q}$. In the long-term limit, $t \gg \tau_u$, the inertial effects from the mass $m$ can be neglected, and the resolvent reduces to our result in Eq.~\eqref{eq:G}, $G_Q = \lim_{t\to\infty}G_u/m$. Furthermore, the MSD \eqref{eq:MSD_passive_2} also coincides with the long-term anomalous diffusion limit of the underdamped equation. Thus, Eq.~\eqref{eq:fLe_Q} describes the overdamped dynamics of the system by assuming that at $t=0$ the stationary state has already been reached.
\end{widetext}

\begin{widetext}
\section{\label{app:msd_act_rot} Derivation of the active rotational MSD\textbf{---}} 
Replacing the autocorrelation of the self-propulsion direction, given by \cref{eq:correlation_rot}, in the expression for the total MSD \eqref{eq:MSD_active} and changing the integration domain as in Eq.~\eqref{eq:MSD_passive_2}, we obtain
\begin{equation}
\begin{aligned}[b]
\langle (\boldsymbol{r}-\boldsymbol{r}_0)^2 \rangle_a 
& = \upsilon^2 \int_0^td\tau_1\int_0^td\tau_2\langle\boldsymbol{\hat{n}}(\tau_1)\cdot\boldsymbol{\hat{n}}(\tau_2)\rangle  \\
& = \upsilon^2 \int_0^td\tau_1\int_0^td\tau_2\;e^{-\frac{D_{R}}{2}|\tau_1-\tau_2|^{\alpha_R}} \\
& = 2\upsilon^2 \int_0^td\tau_1\int_0^{\tau_1}d\tau_2\;e^{-\frac{D_{R}}{2} (\tau_1-\tau_2)^{\alpha_R}} \\
& = 2\upsilon^2 \int_0^td\tau_1\int_0^{\tau_1}d\tau\;e^{-\frac{D_{R}}{2} \tau^{\alpha_R}},
\end{aligned}
\label{eq:MSD_active_rot1}
\end{equation}
where we made the change of variables $\tau=\tau_1-\tau_2$.

Now, using the power-series representation of the exponential and integrating, we obtain
\begin{equation}
\begin{aligned}[b]
\langle (\boldsymbol{r}-\boldsymbol{r}_0)^2 \rangle_a
 & = 2\upsilon^2 \int_0^td\tau_1\int_0^{\tau_1}d\tau\;\sum_{k=0}^\infty\frac{\left(-{\frac{D_{R}}{2}\tau^{\alpha_R}}\right)^{k}}{k!} \\
& = 2 \upsilon^2 t^2 \sum_{k=0}^\infty\frac{\left(-{\frac{D_{R}}{2}t^{\alpha_R}}\right)^{k}}{k!(k\alpha_R+1)(k\alpha_R+2)} \\
& = \frac{2\upsilon^2t^2}{\alpha_R} \left[\sum_{k=0}^\infty\frac{\left(-{\frac{D_{R}}{2}t^{\alpha_R}}\right)^{k}}{k!(k+1/\alpha_R)}  - \sum_{k=0}^\infty\frac{\left(-{\frac{D_{R}}{2}t^{\alpha_R}}\right)^{k}}{k!(k+2/\alpha_R)}\right]. \\
\end{aligned}
\label{eq:MSD_active_rot2}
\end{equation}

The last line in Eq.~\eqref{eq:MSD_active_rot2} can be simplified using the power-series representation of the lower incomplete Gamma function,
\begin{equation}
\gamma(z,a) = \int_0^at^{z-1}e^{-t}dt = a^z\sum_{k=0}^\infty\frac{(-a)^k}{k!(k+z)},
\end{equation}
to obtain
\begin{equation}
\langle (\boldsymbol{r}-\boldsymbol{r}_0)^2 \rangle_a =
    \frac{2\upsilon^2}{\alpha_R(D_{R}/2)^{\frac{2}{\alpha_R}}}\left[ \gamma\left(\frac{1}{\alpha_R},\frac{D_{R}}{2}t^{\alpha_R}\right)\left(\frac{D_{R}}{2}\right)^{\frac{1}{\alpha_R}}t- \gamma\left(\frac{2}{\alpha_R},\frac{D_{R}}{2}t^{\alpha_R}\right)\right].
\end{equation}

\end{widetext}
\end{document}